\documentclass[final,1p,times]{elsarticle} 
\pdfoutput=1
\usepackage{graphicx} 
\usepackage{amssymb} 
\usepackage{amsthm} 
\usepackage{lineno} 
\usepackage{hyperref}

\newcommand{\jpsi}{{\rm{J}/\psi}}

\newcommand{\da}{{d+{\rm{A}}}}
\newcommand{\pa}{{p+{\rm{A}}}}
\newcommand{\dau}{{d+{\rm{Au}}}}
\newcommand{\auau}{{\rm Au}+{\rm Au}}

\newcommand{\pp}{{p+p}}
\newcommand{\pbpb}{{\rm Pb}+ {\rm Pb}}
\newcommand{\ncol}{{N_{\rm col}}}

\newcommand{\raa}{{R_{\rm AA}}}
\newcommand{\raab}{{R_{\rm AA}^{\rm B}}}
\newcommand{\raad}{{R_{\rm AA}^{\rm D}}}

\renewcommand{\aa}{{\rm{A}+\rm{A}}}
\newcommand{\pt}{{p_{\rm T}}}

\journal{Nuclear Physics A} 
\begin{document} 

\begin{frontmatter} 

\title{Early times and thermalization in heavy ion collisions: a summary of experimental results for photons, light vector mesons, open and hidden heavy flavors}

\author{Hugo Pereira Da Costa}

\address{IRFU/SPhN, CEA Saclay, F-91191, Gif-sur-Yvette, France}

\begin{abstract} 
This contribution summarizes the main experimental results presented at the 2009 Quark Matter conference concerning single and dilepton production in proton and heavy ion collisions at high energy. The dilepton invariant mass spectrum has been measured over a range that extends from the $\pi^0$ mass to the $\Upsilon$ mass, and for various collision energies at SPS, Fermilab, Hera and RHIC. This paper focuses on the various contributions (photons, low mass vector mesons, open and hidden heavy flavors) to this spectrum and discuss their implications on our understanding of the matter formed in heavy ion collisions.
\end{abstract} 

\end{frontmatter} 

\section{Introduction}

Single and dilepton probes in heavy ion collisions are of particular interest since such probes, once produced, are largely unaffected by the surrounding QCD medium. They carry valuable information on the particle from which they originate and allow one to assess the properties of the medium formed in the early instants of the collision. The following contributions to the dilepton invariant mass spectrum are discussed 
here, 
together with what one might learn from their measurement about the properties of the medium formed in the collision:
\begin{itemize}
\item Low mass dileptons originating from vector meson leptonic decay ($\rho$, $\phi$ and $\omega$) provide insight on the properties of these mesons in the high temperature expanding fireball produced immediately after the collision, where chiral symmetry may be (at least partially) restored~\cite{Pisarki, Brown, Rapp};
\item A significant fraction of the virtual and direct photons produced at low $\pt$ ($\pt<1$~GeV/c) in heavy ion collisions originates from the thermal black-body radiation of the created fireball~\cite{Stankus, Turbide}. Measuring these photons therefore allows one to quantify the temperature of the fireball;
\item Open heavy flavors, because of their high mass, allow one to study in-medium energy loss mechanisms in addition to what can be learned from light quarks~\cite{Baier, Gyulassy};
\item Heavy quarkonia are of interest because of additional mechanisms that are predicted to occur in the presence of a QGP and that would affect the production of these bound states~\cite{Matsui, Andronic, Thews}. 
\end{itemize}

\section{Low mass vector mesons}

\begin{figure}[ht]
\centering
\begin{tabular}{cc}
\hspace*{-5mm}\includegraphics[height=6cm]{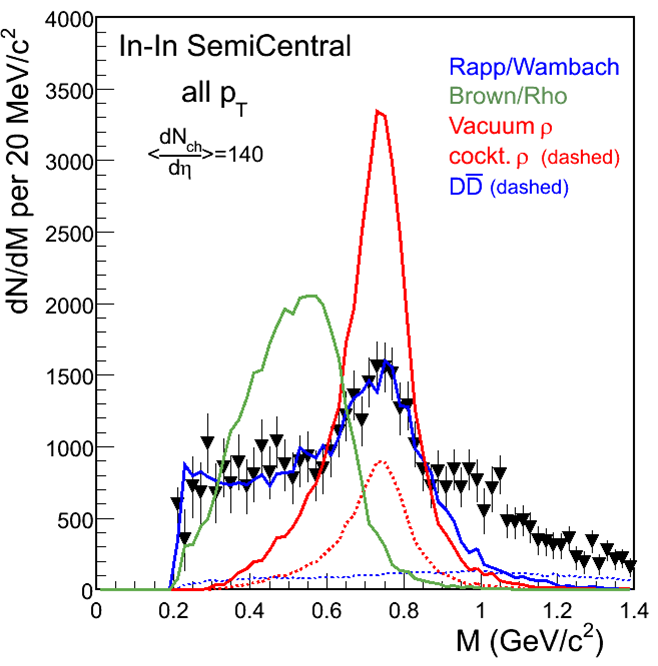}&
\hspace*{-5mm}\includegraphics[height=6cm]{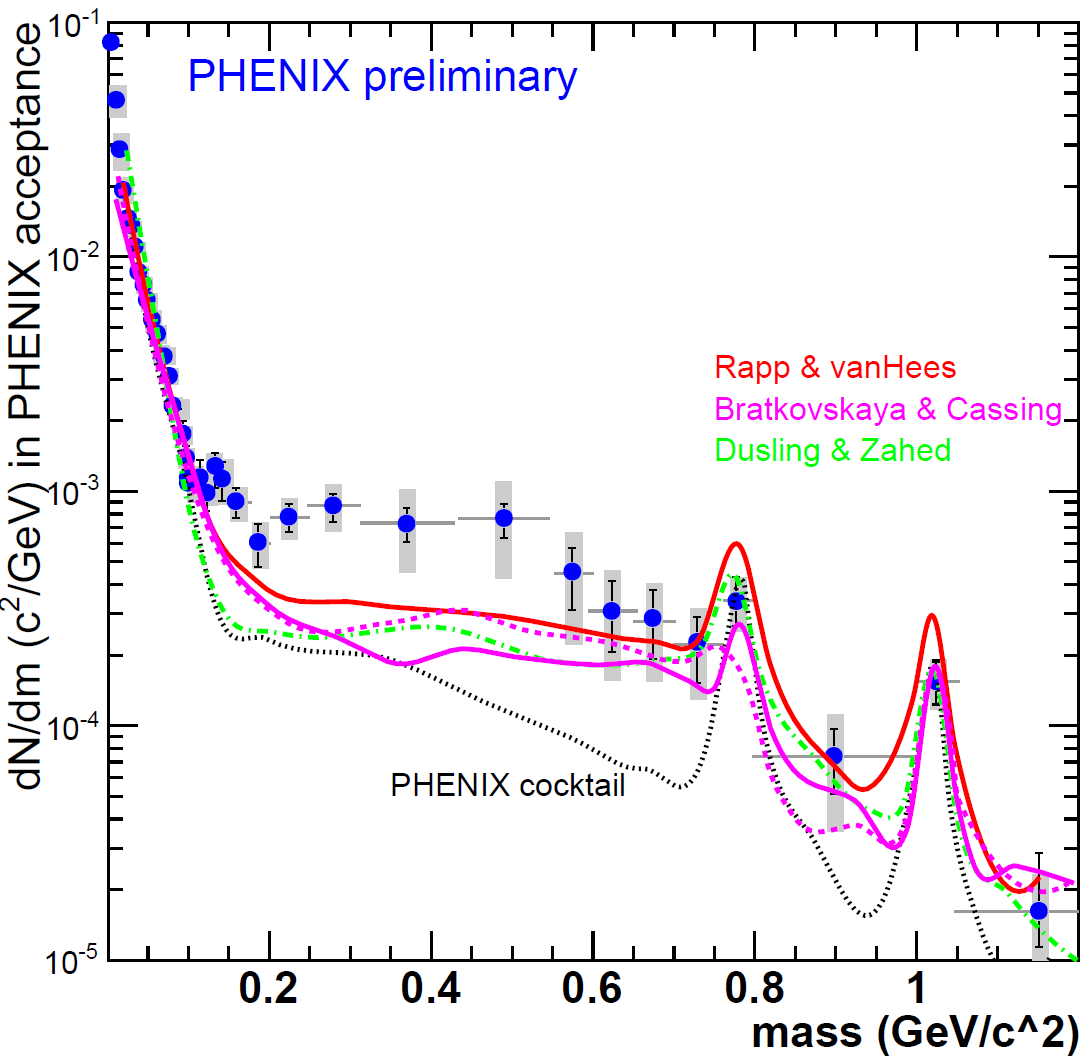}\\
\end{tabular}
\caption[]{\label{low_mass_vector_mesons}
Left: dimuon invariant mass distribution in the $\rho$ mass region in In+In semi-central collisions measured by NA60 at SPS. Right: low mass dielectron invariant mass distribution in Au+Au collisions measured by PHENIX at RHIC.}
\end{figure}

Fig.~\ref{low_mass_vector_mesons} (left) shows the correlated dimuons invariant mass distribution at the $\rho$ vacuum mass, measured by the NA60 experiment in semi-central In+In collisions~\cite{NA60_rho}. The $\rho$ mass peak differs significantly from the expected vacuum $\rho$ and can be reasonably well described on the low mass side by the model presented in~\cite{Rapp,NA60_rapp}. This model includes a detailed description of the baryonic matter created in the collision below the formation temperature of a QGP, $T_c$. Interactions with this baryonic matter are responsible for a broadening of the $\rho$ (but no modification of its mass) when approaching chiral symmetry restoration near $T_{c}$. 

A measurement of dilepton invariant mass distributions in the same mass region has been carried out by the PHENIX collaboration at RHIC in Au+Au collisions at $\sqrt{s_{NN}} = 200$~GeV~\cite{low_mass_aa_rhic}. An excess over expected background sources is observed between $0.1$ and $0.6$~GeV/c$^2$ which cannot be described by models similar to the one above~\cite{NA60_rapp}, although such models work reasonably well for larger masses  (Fig.~\ref{low_mass_vector_mesons}, right). This low mass excess is larger for low $\pt$ dileptons. A possible contribution to this excess, which has not been accounted for in the calculations above, might come from quark-gluon scattering into a quark and a virtual photon ($qg\rightarrow q\gamma^*$). A similar calculation valid for the direct photon production yields at RHIC has been carried out in \cite{Direct_photon_rapp}, which 
accounts for $q+g$ scattering using a complete leading-order QGP emission rate~\cite{Direct_photon_arnold}.
The predicted integrated magnitude of this contribution is about one third of the hadron gas thermal radiation contribution. Applying this to the virtual photon case might explain part of the excess observed at RHIC, but a detailed calculation is still to be carried out.

\section{Direct photons}
\begin{figure}[ht]
\centering
\begin{tabular}{cc}
\includegraphics[height=5.2cm]{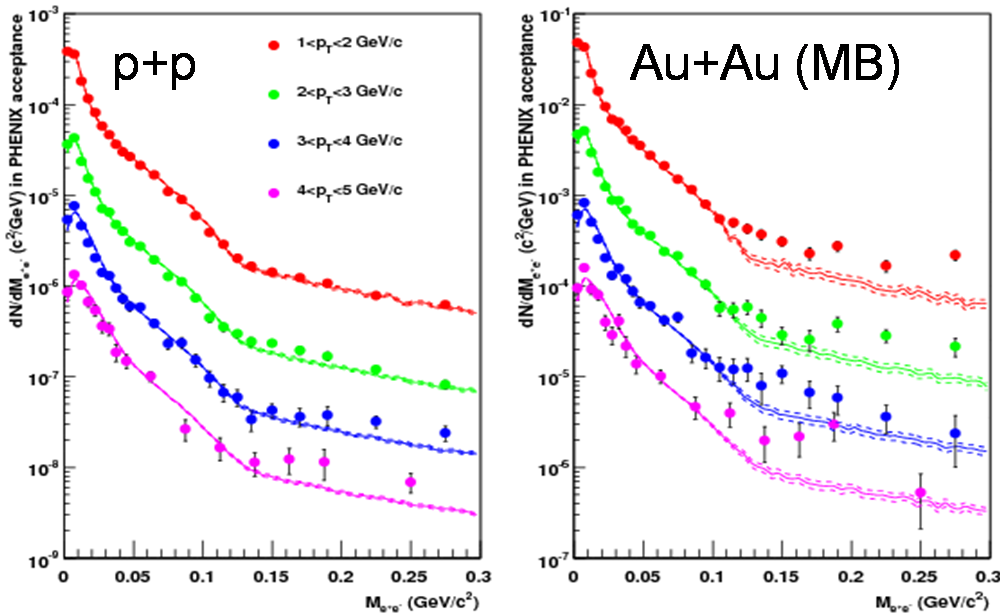}\vspace*{-1mm}&
\includegraphics[height=5.2cm]{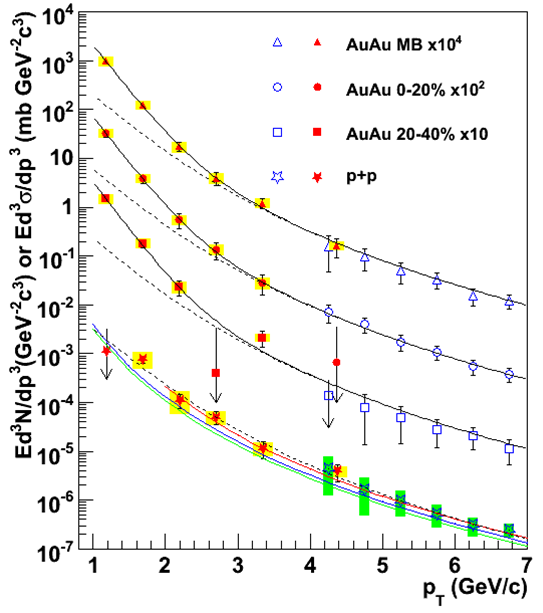}\vspace*{-1mm}
\end{tabular}
\caption[]{Left and center: dilepton invariant mass distribution as a function of mass for different $\pt$ bins, in $\pp$ collisions (left) and Au+Au minimum bias collisions (center). Data are compared to expected background sources to derive a possible virtual photon excess. Right: calculated direct photon yield as a function of $\pt$ in different centrality bins, compared to binary scaled $\pp$ yields.}
\label{direct_photons}
\end{figure}

Direct photon production yields (as a function of $\pt$) can be derived from the dilepton invariant mass spectrum using the following steps~\cite{Akiba}: 1) consider the excess of dileptons over expected hadronic sources in the kinematic range $m\in[0.1,0.3]$~GeV/c$^2$ and $\pt>1$~GeV/c, where the contribution of low mass vector mesons should be negligible (Fig.~\ref{direct_photons}, left and center panels), 2) interpret this excess as a direct virtual photon signal (with photons decaying into dielectrons) and 3) extrapolate this signal to an invariant mass $m=0$ to get the corresponding real photon production yield.  

The resulting yields (as a function or $\pt$) are compared to yields obtained in $\pp$ collisions scaled by $\ncol$, the number of nucleon-nucleon collisions equivalent to one $\aa$ collision in a given centrality bin, and the difference is fitted to extract a time averaged (over the medium expansion history) {\em black body} radiation temperature (Fig.~\ref{direct_photons}, right). For central Au+Au collisions at RHIC energy, a temperature of 221$\pm$23~MeV is obtained~\cite{Direct_photon_phenix}. These thermal photon yields can also be compared to various theoretical models in order to derive a medium {\em initial} temperature, by making assumptions on how this medium expands and cools down over time~\cite{Direct_photon_enterria}. Depending on how long it takes for the system to thermalize, an initial temperature between $300$ and $600$~MeV is obtained. As one might expect, later thermalization times led to smaller initial temperatures. Similar fits applied to the $\pbpb$ WA98 direct photon measurements~\cite{Direct_photon_wa98} give an initial temperature of about $200$~MeV~\cite{Direct_photon_wa98_theo}.
 
\section{Open heavy flavor}

\begin{figure}[t]
\centering
\vspace*{-15mm}
\begin{tabular}{cc}
\includegraphics[height=4.5cm]{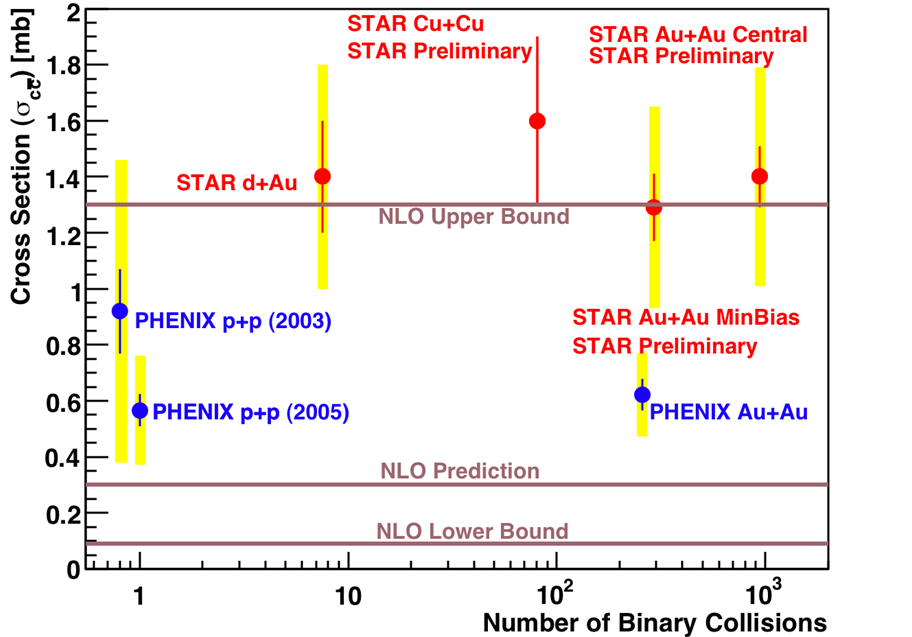}&
\includegraphics[height=4.5cm]{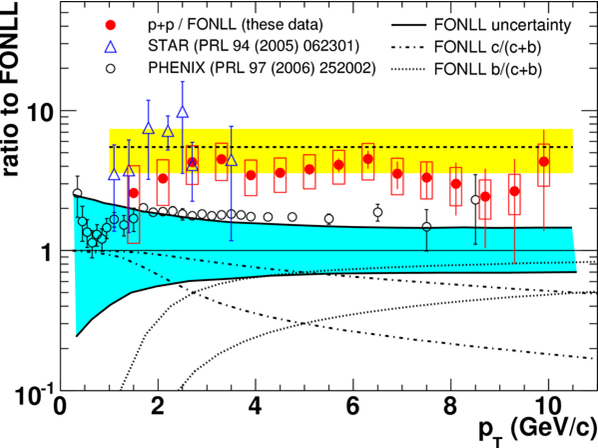}\\
\end{tabular}
\caption[]{Left: total heavy-flavor production cross-section as a function of $\ncol$ measured by PHENIX and STAR at RHIC in $\pp$, $\da$ and $\aa$ collisions. Right: ratio between the heavy-flavor differential production cross-section as a function of decay electron $\pt$ measured by PHENIX and STAR in $\pp$ collisions and a FONLL calculation.}
\label{heavy_flavor}
\end{figure}

There is still a disagreement of about a factor two between the STAR and PHENIX heavy flavor (charm and beauty) total cross-section measurements in $\pp$, $\da$ and $\aa$ collisions at $\sqrt{s}=200$~GeV~\cite{heavy_flavor_star,heavy_flavor_raa_phenix}, as well as between the open charm differential cross-section as a function of $\pt$~\cite{heavy_flavor_star} (Fig.~\ref{heavy_flavor}). The main differences between the two experiments are 1) the amount of material in the detector acceptance
2) the rapidity and $\pt$ range of the measured electrons used for heavy flavor identification. 
Efforts are underway in both collaborations to better understand existing measurements and provide new independent measurements in order to address this discrepancy:
\begin{itemize}
\item The PHENIX collaboration is working on refining its understanding of the electron cocktail which is subtracted from the raw single electron spectrum to derive the heavy-flavor signal, and now accounts for the contribution of electrons coming from $\jpsi$, $\Upsilon$ and Drell-Yan~\cite{Dion}. PHENIX also measured the total D+B production cross-section in a largely independent way by estimating all the contributions to the dielectron invariant mass spectrum (as opposed to the single electron spectrum) using data-driven simulations~\cite{low_mass_pp_rhic}. Finally PHENIX reported on a first study of electron-muon correlations to measure $D\overline{D}$ production in a way that is largely free of background~\cite{Tatia};
\item The STAR collaboration has removed its central silicon detector in order to reduce the amount of material in the spectrometer and the corresponding photo-conversion background contribution to the raw single electron spectrum. It also measured the production of low $\pt$ $D$ mesons using their decay into a $K,\pi$ pair, and using single muons~\cite{D_star, D_star_mu};
\end{itemize}

\begin{figure}[ht]
\vspace*{-8mm}
\centering
\begin{tabular}{cc}
\includegraphics[height=5.0cm]{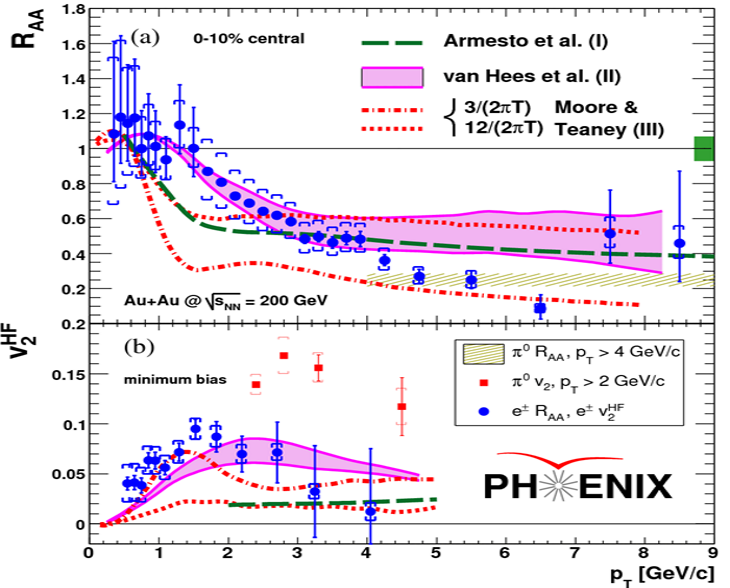}&
\includegraphics[height=4.2cm]{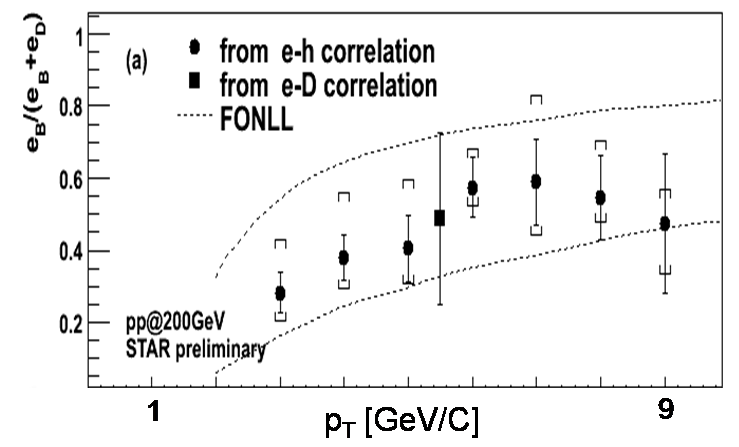}
\end{tabular}
\caption[]{Left: Heavy-flavor electron $\raa$ and elliptic flow measured by PHENIX in Au+Au collisions at RHIC; right: B/D+B production ratio as a function of $\pt$ in $\pp$ collisions measured by STAR, compared to FONLL calculations.}
\label{heavy_flavor_aa}
\end{figure}

In $\aa$ collisions, the measurement of the heavy flavor nuclear modification factor $\raa$ agrees between the two collaborations~\cite{heavy_flavor_star,heavy_flavor_raa_phenix}. The heavy flavor production at high $\pt$ ($\pt>3$~GeV/c) exhibits a large suppression with respect to binary scaled cross-sections in $\pp$ (Fig.~\ref{heavy_flavor_aa}, top-left). This indicates that high $\pt$ heavy quarks lose a significant fraction of their energy when traversing the medium created during the collision, and poses a challenge to theoretical models, since heavy quarks, due to their high mass, are expected to loose less energy (via gluon radiation) than light quarks~\cite{Kharzeev}. Additionally, a large elliptic flow $v_2$ is observed for intermediate $\pt$ heavy quarks ($1<\pt<3$~GeV/c) in $\auau$ minimum bias collisions (Fig.~\ref{heavy_flavor_aa}, bottom-left), indicating that intermediate $\pt$ heavy quarks are rapidly thermalized. These two observations are interpreted as an evidence for a strong coupling between the heavy quarks and the medium produced during the collision. No consensus amongst theorists has been achieved to date concerning the underlying mechanism responsible for this strong coupling (see e.g.~\cite{Armesto,VanHees,Moore}. 

Current single lepton measurements do not allow for a separation of charm and beauty in a model independent way. However, separate measurements have been performed to determine the relative contribution of charm and beauty to total heavy flavor yields. These are either direct measurements (using the hadronic decay of D mesons), or indirect measurements (e.g. by studying the correlation of opposite sign electron-hadron pairs in the final state to separate the contributions of D and B semi-leptonic decays). In $\pp$ collisions, the resulting B/(B+D) ratios agree well between STAR and PHENIX~\cite{Dunlop}. They are consistent with a Fixed Order Next to Leading Log (FONLL) calculation~\cite{Cacciary} (Fig.~\ref{heavy_flavor_aa}, right). 

Measuring the total heavy flavor $\raa$ and the B/(B+D) ratio in $\pp$ collisions allows one to uniquely relate the $\raa$ of B and D mesons: smaller values of $\raad$ bring $\raab$ closer to unity. The (negative) slope of the relation between the two is driven by the D/B ratio measured in $\pp$ collisions whereas its magnitude is controlled by the total heavy flavor $\raa$. The main conclusion of such an analysis~\cite{Dunlop} is that even in the unlikely case where high $\pt$ charm quarks are entirely suppressed in $\aa$ collisions, a significant suppression of high $\pt$ $b$ quarks is still needed to explain the total heavy flavor $\raa$ measured at RHIC. This poses an even greater challenge to theoretical models than the charm $\raa$, since $b$ quarks are significantly heavier than $c$ quarks.

More information will be gained on this matter by measuring charm and beauty separately in $\aa$ collisions. Both STAR and PHENIX are undergoing silicon vertex detector upgrades for the central tracking that should allow direct measurement of D and B mesons.

\section{Heavy Quarkonia}

Heavy quarkonia have been studied extensively at the SPS and the RHIC since they are predicted to melt, via QCD Debye screening, in the presence of a Quark-Gluon Plasma~\cite{Matsui}. Recently, focus has been given to understanding both the heavy quarkonia production mechanism in $\pp$ collisions and the cold nuclear matter effects which affect the production of heavy quarkonia when colliding two nuclei without the formation of a QGP. 

\begin{figure}[ht]
\centering
\vspace*{-3mm}
\begin{tabular}{cc}
\includegraphics[height=4.8cm]{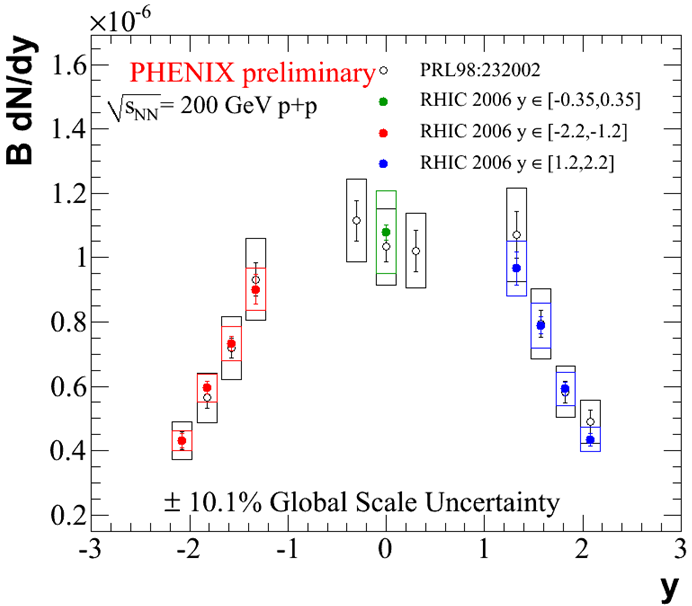}&
\includegraphics[height=4.5cm]{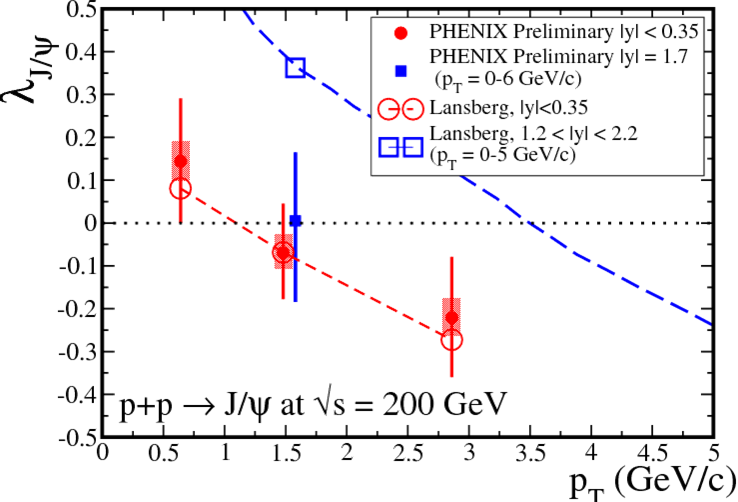}
\end{tabular}
\caption[]{Left: $\jpsi$ production yield as a function of $\jpsi$ rapidity measured in $\pp$ collisions at RHIC. Right: $\jpsi$ polarization measured in the helicity frame by PHENIX in $\pp$ collisions.}
\label{jpsi_pp_rhic}
\end{figure}

Heavy quarkonia production yields in $\pp$ collisions serve as a reference to study medium effects in $\pa$, $\da$ and $\aa$ collisions and help in understanding how these bound states are produced. Fig.~\ref{jpsi_pp_rhic} (left) shows the $\jpsi$ production invariant yields as a function of rapidity measured in $\pp$ collisions at RHIC by PHENIX using the 2006 high statistics $\pp$ data sample~\cite{Cesar}. These yields can be compared to calculations that assume different underlying production mechanisms, however both statistical and systematic uncertainties are still too large to uniquely identify the correct mechanism at play. Another way to address the production mechanism is to measure the $\jpsi$ polarization since models have very different predictions for this observable. Fig.~\ref{jpsi_pp_rhic} (right) shows the $\jpsi$ polarization measured in the helicity frame by PHENIX in $\pp$ collisions at mid and forward rapidity~\cite{Cesar}. The model shown on the figure (a refined version of the Color Singlet Model \cite{lansberg}) reproduces reasonably well the data at mid-rapidity but misses the measurement at forward rapidity. Similarly, all available measurements on $\jpsi$ polarization have been collected, {\em rotated} so that they are all evaluated in the same reference frame (here the Collin-Sopper frame~\cite{collins}) and represented as a function of the $\jpsi$ momentum. A global trend is observed that is largely independent of the collision energy but lacks a theoretical explanation~\cite{jpsi_polarization}.

Cold nuclear matter effects must be carefully evaluated and properly accounted for when considering yield modifications observed in $\aa$ collisions before quantifying the effects of a QGP. They include: modification of the parton distribution functions (pdf) in the nucleus (notably shadowing or gluon saturation at low $x_{\rm Bj}$, anti-shadowing at large $x_{\rm Bj}$); nuclear absorption/dissociation; initial state energy loss and the Cronin effect. The general approach used up to now to quantify the cold nuclear matter effects~\cite{dau_rhic} is to choose a set of modified pdfs, add some effective absorption (or break-up) cross-section to account for the other possible effects, derive the resulting expected heavy quarkonia production yield, and fit this expected yield to the $\pa$ or $\da$ available measurements, leaving the effective break-up cross-section as a free parameter. These effects are then extrapolated to $\aa$ collisions and compared to the data.

\begin{figure}[ht]
\centering
\vspace*{-6mm}
\begin{tabular}{cc}
\includegraphics[height=5cm]{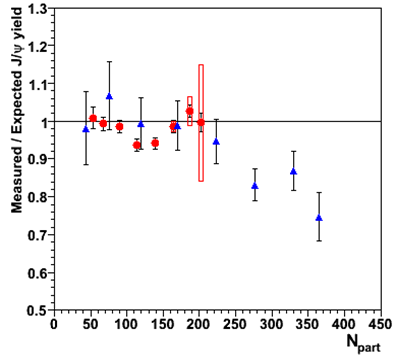}\vspace*{-1mm}&
\includegraphics[height=5cm]{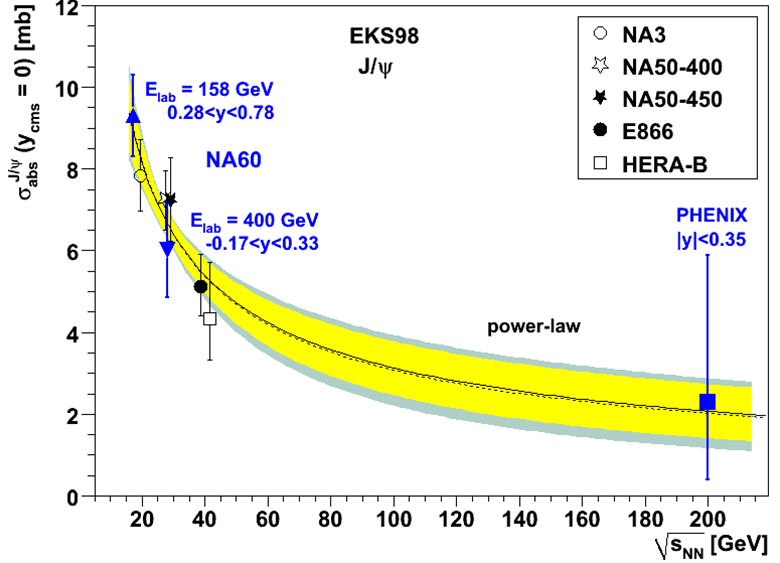}\vspace*{-1mm}
\end{tabular}
\caption[]{Left: $\jpsi$ $\raa$ at SPS after removal of CNM effects measured by NA60. Right: $\jpsi$ effective break-up cross-section as a function of collision energy in $\da$ or $\pa$ collisions.}
\label{jpsi_raa_sps}
\end{figure}

At the SPS, an updated break-up cross-section has been estimated that properly accounts for the fact that the gluon $x$ domain covered by the experiments corresponds to the anti-shadowing region of modified pdfs, for which the gluon content is enhanced with respect to the bare nucleon case (see e.g.~\cite{EPS09}). Consequently, the new cross-section derived from $\pa$ data is significantly larger than the previously published value. When extrapolated to In+In, the $\jpsi$ suppression factor estimated from cold nuclear matter effects matches the data rather well and leaves little room for any additional anomalous suppression~\cite{Scompari} (Fig.~\ref{jpsi_raa_sps}, left).

At RHIC, updated break-up cross-sections have been derived from the new 2009 $\dau$ data sample which is about 30 times as large as the one used for previous published results~\cite{Cesar}. These cross-sections must still be extrapolated to Au+Au collisions in order to quantify any additional anomalous suppression due to the possible formation of a QGP.

A systematic survey of the effective charmonia break-up cross-section has been performed that collects results from SPS, HERA, Fermilab and RHIC~\cite{jpsi_breakup}. When plotted as a function of the collision energy a common (exponentially decreasing) trend is observed although this trend has no theoretical interpretation yet (Fig.~\ref{jpsi_raa_sps}, right). When represented as a function of rapidity, and disregarding the collision energy, the effective break-up cross-section also exhibits a somewhat universal trend, that cannot be easily explained in terms of the effects listed above. Note that similar surveys have been performed in the past that led to different conclusions, namely that the current data are consistent with no energy dependency~\cite{Tram}.

\begin{figure}[ht]
\centering
\begin{tabular}{ccc}
\hspace*{-5mm}\includegraphics[height=3.7cm]{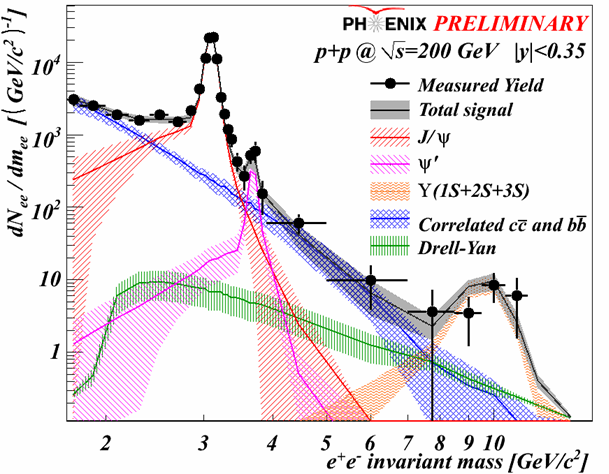}&
\hspace*{-3mm}\includegraphics[height=3.7cm]{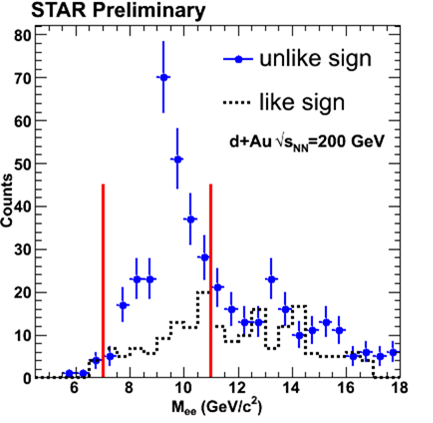}&
\hspace*{-4mm}\includegraphics[height=3.7cm]{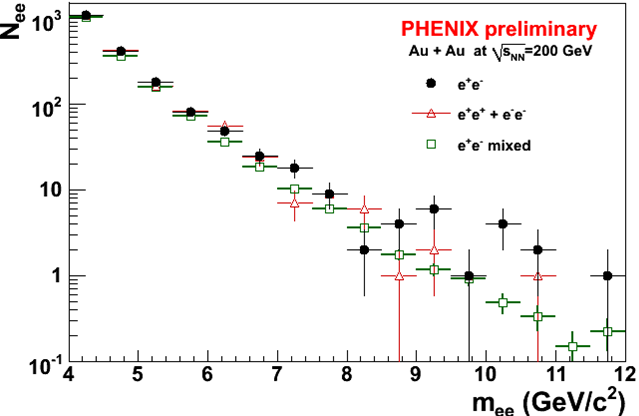}
\end{tabular}
\caption[]{dielectron invariant mass distributions at high mass in $\pp$ (left), $\dau$ (center) and Au+Au collisions, measured by PHENIX and STAR at RHIC.}
\label{upsilon}
\end{figure}

The first $\Upsilon$ measurements have become available at RHIC (with limited statistics) in $\pp$, $\dau$ and $\auau$ collisions (Fig.~\ref{upsilon}). 
Due to limited statistics, it is difficult to disentangle the $\Upsilon$ signal and the underlying correlated background sources (from Drell-Yan and open beauty). One can either ignore these contributions and derive e.g. nuclear modification factors for inclusive high-mass dileptons, or estimate them from simulations and use the corresponding uncertainty as a systematic error. In $\pp$ collisions a total $\Upsilon$ production cross-section $BR d\sigma/dy (|y|<0.35) = 114^{+46}_{-45}$~pb is measured~\cite{Cesar}; in $\dau$ collisions a nuclear modification factor consistent with unity is observed~\cite{Liu} while in $\auau$ collisions this nuclear modification factor is smaller that 0.64 at 90~\% confidence level~\cite{Levy}, meaning that inclusive high mass dileptons are significantly suppressed by the medium formed in $\auau$ collisions at $\sqrt{s_{\rm NN}}=200$~GeV.

\section{Conclusion}
In short:

\begin{itemize}
\item Low mass vector mesons exhibit strong shape modifications with respect to their vacuum properties, that can be well described at SPS but not at RHIC possibly because some contributions to the dilepton spectrum have not been properly accounted for;
\item Virtual photons can be used in addition to direct photon measurements to assess the medium temperature averaged over its expansion time and derive its initial temperature;
\item A significant suppression of $b$ quarks is necessary to describe the observed heavy flavor $\raa$ in a way that is consistent with the B/B+D ratio measured in $\pp$ collisions;
\item $\jpsi$ production in heavy-ion collisions is a puzzle. The situation is more complex than the original picture, due to our poor knowledge of its production mechanism in $\pp$ collisions and to the existence of many cold nuclear matter effects which significantly modify this production even in the absence of a QGP. Efforts are being made to better understand the above so that one can 
quantify the {\em hot}, abnormal effects at both SPS and RHIC. Notably, it appears that the suppression measured at SPS in In-In collisions can be entirely described in terms of such cold nuclear matter effects. 
\end{itemize}

\end{document}